\documentclass[pdftex,11pt,a4paper]{article}
\usepackage[pdftex]{color,graphicx}
\usepackage[margin=0.8in]{geometry}
\usepackage[english]{babel}
\usepackage[utf8]{inputenc}
\usepackage{subcaption}
\usepackage{parskip,tocbibind}
\usepackage{amsmath,amssymb,amsthm,braket,enumitem}
\usepackage{commath}
\usepackage{hyperref}
\hypersetup{
    colorlinks=true,
    linkcolor=blue,
    filecolor=magenta,      
    urlcolor=blue,
    citecolor=blue
}
\usepackage{cleveref}
\usepackage{float}
\usepackage{cite}
\usepackage{authblk}

\renewcommand{\d}{\textnormal{d}}
\newcommand{\smallfrac}{\genfrac{}{}{}1}

\title{Exact analytical solution of a novel modified nonlinear Schr\"{o}dinger equation: solitary quantum waves on a lattice}
\author[1]{Jingxi Luo}
\affil[1]{School of Mathematics, University of Birmingham, Birmingham, B15 2TT, United Kingdom. j.luo.5@bham.ac.uk}
\date{}

\begin{document}
\pagestyle{plain}
\setlength{\baselineskip}{16pt}
\maketitle

\begin{abstract}
A novel modified nonlinear Schr\"{o}dinger equation is presented. Through a travelling wave ansatz, the equation is solved exactly and analytically. The soliton solution is characterised in terms of waveform and wave speed, and the dependence of these properties upon parameters in the equation is detailed. It is discovered that some parameter settings yield unique waveforms while others yield degeneracy, with two distinct waveforms per set of parameter values. The uni-waveform and bi-waveform regions of parameter space are identified. Finally, the equation is shown to be a model for the propagation of a quantum mechanical exciton, such as an electron, through a collectively-oscillating plane lattice with which the exciton interacts. The physical implications of the soliton solution are discussed. 
\end{abstract}


\textbf{This is the peer reviewed version of the following article: [Luo, J. (2020). Exact analytical solution of a novel modified nonlinear Schr\"{o}dinger equation: Solitary quantum waves on a lattice. \textit{Studies in Applied Mathematics}.], which has been published in final form at \url{https://doi.org/10.1111/sapm.12355}. This article may be used for non-commercial purposes in accordance with \href{https://authorservices.wiley.com/author-resources/Journal-Authors/licensing/self-archiving.html}{Wiley Terms and Conditions for Use of Self-Archived Versions}.}

    \section{Introduction}

The nonlinear Schr\"{o}dinger equation (NLSE) and its generalisations have interested researchers for many years, due to the richness of their mathematical structure and the breadth of their applications to fields such as electronics, optics, plasma physics, molecular biophysics, \emph{et cetera} \cite{shabat1972exact,hirota1973exact,kevrekidis2016solitons,davydov1979solitons,luo2018directed}. A natural generalisation is raising the number of spatial dimensions. Exact analytical solutions to the 2D NLSE are known, and relate to the propagation of acoustic disturbances in a static cosmological background \cite{Seadawy2012,kates1988two}. The 3D version, also known as the Gross–Pitaevskii equation, has been studied for the existence and blow-up properties of its solutions, and is a model for Bose-Einstein condensates \cite{holmer2007,holmer2010,duyckaerts2008scattering,dalfovo1999theory,garcia2003quasi}. An $N$-dimensional variant where the cubic nonlinearity is replaced by a $p$\textsuperscript{th}-power nonlinearity, with $ p > 1 + \smallfrac{4}{N} $, has been analysed, its blow-up criteria established \cite{duyckaerts2015going}. 

In other developments, a modified NLSE has been proposed as a continuum model of electron self-trapping in an $N$-dimensional lattice, and its stationary solutions have been found \cite{brizhik2001electron,brizhik2003static,Luo2017}. Non-stationary, \textit{soliton} solutions exist for many generalisations or modifications of the NLSE, including ones with higher-power nonlinear terms, trigonometric nonlinear terms or extra derivatives, and studies of these solutions have influenced modern technologies such as optical fibres \cite{hasegawa1973transmission1,biswas2008femtosecond,chen2010formation,hosseininia2018wavelet}, with some solutions having analytical expressions \cite{potasek1991exact,bulut2018optical,benia2019exact,gao2020instability}. However, exact soliton solutions to modified NLSEs remain rare. 

In this work, we investigate a novel modified NLSE in two spatial dimensions, and construct an exact analytical solution representing a soliton. We show that the equation has physical origins in the system of an exciton coupled to a collectively oscillating plane lattice, and hence that the soliton solution represents a propagating exciton wavefunction in the lattice. We characterise the soliton's shape and velocity, and detail how they depend on parameters of the physical system. Though we focus on a 2D equation, the method developed herein is readily applicable to higher-dimensional generalisations. 

    \section{The equation and soliton solutions}

We consider the equation,
    \begin{align}
    & i \partial_t \psi + \partial_{xx} \psi + \sigma \partial_{yy} \psi + \kappa | \psi |^2 \psi - \lambda ( \partial_{xx} | \psi |^2 ) \psi = 0, \label{eq-GNLSE}
    \end{align}
for a complex-valued function $ \psi $ of $ x \in \mathbb{R} $, $ y \in (-\smallfrac{1}{2},\smallfrac{1}{2}) $, $ t > 0 $, with real constants $ \kappa > 0, \lambda > 0 $, and $ \sigma \neq 0 $. We seek non-trivial solutions with vanishing value and derivatives at $ x \rightarrow \pm \infty $, which are normalised:
    \begin{align}
    \int_{-1/2}^{1/2} \int_{-\infty}^{\infty} | \psi (x,y,t) |^2 ~\d x ~\d y = 1. \label{eq-GNLSE-norm}
    \end{align}
The travelling wave ansatz, 
    \begin{align}
    \psi (x,y,t) = f ( \xi ) \exp( i \eta ), \quad \textnormal{where } \xi = x + ay - [ 2 + \sigma b (1+a) ] t, \quad \eta = x + by - Et,
    \end{align}
where $ f $ is a real smooth function and $ a,b,E $ are real constants, transforms \eqref{eq-GNLSE} into
    \begin{align}
    - ( 1 + \sigma b^2 - E ) f + ( 1 + \sigma a ) f'' + \kappa f^3 - \lambda (f^2)'' f = 0, \label{eq-GNLSE-2}
    \end{align}
with $'$ denoting differentiation with respect to $ \xi $. 

\subsection*{Case I. $ 1 + \sigma a = 0 $.}

With $ 1 + \sigma a = 0 $, \eqref{eq-GNLSE-2} is significantly simplified. Requiring $f \not \equiv 0$, we deduce $ - ( 1 + \sigma b^2 - E ) + \kappa f^2 - \lambda (f^2)'' = 0 $, which is a linear ODE for $f^2$ with the solution $ f^2 = (1 + \sigma b^2 - E) / \kappa + B \exp ( \xi \sqrt{ \kappa / \lambda } ) + C \exp ( - \xi \sqrt{ \kappa / \lambda } ) $, for some arbitrary constants $ B,C $. It is therefore impossible for a non-trivial $f$ to satisfy vanishing boundary conditions, and so we must require $ a \neq - 1 / \sigma $.

\subsection*{Case II. $ 1 + \sigma a > 0 $.}

With $ 1 + \sigma a > 0 $, we use $ ( f^2 )'' \equiv 2ff'' + 2(f')^2 $ to rewrite \eqref{eq-GNLSE-2} as
    \begin{align}
    - E_0 f + ( 1 - 2 \lambda_0 f^2 ) f''  + \kappa_0 f^3 - 2 \lambda_0 (f')^2 f = 0, \label{eq-GNLSE-caseII}
    \end{align}
where
    \begin{align}
    E_0 = \frac{1 + \sigma b^2 - E}{1 + \sigma a}, \quad \lambda_0 = \frac{\lambda}{1 + \sigma a} > 0, \quad \kappa_0 = \frac{\kappa}{1 + \sigma a} > 0. \label{eq-GNLSE-caseII-consts}
    \end{align}
Noticing that $ f' \d f' / \d f \equiv f'' $ (assuming $f'$ is a well-defined function of $f$), and therefore $ ( 1 - 2 \lambda_0 f^2 ) f''  - 2 \lambda_0 (f')^2 f \equiv \frac{\d}{\d f} [\frac{1}{2} ( 1 - 2 \lambda_0 f^2 ) (f')^2 ] $, we integrate \eqref{eq-GNLSE-caseII} with respect to $f$ to obtain
    \begin{align}
    ( 1 - 2 \lambda_0 f^2 ) (f')^2 = E_0 f^2 - \frac{1}{2} \kappa_0 f^4 + C, \label{eq-GNLSE-caseII-2}
    \end{align}
where $C = 0$ since $ f $ has vanishing value and derivatives at infinity.

Suppose $ 1 - 2 \lambda_0 f^2 $ could vanish at some value of $f$. Then both sides of \eqref{eq-GNLSE-caseII-2} must vanish at that $f$, from which we derive $ 4 E_0 \lambda_0 / \kappa_0 = 1 $, which is a highly restrictive condition on the parameters. To allow for a variety of travelling wave behaviour depending on parameter values, we reject the restrictive condition and require that $ 1 - 2 \lambda_0 f^2 $ never vanishes. Since $f$ is smooth and $ 1 - 2 \lambda_0 f(\infty)^2 = 1 > 0 $, we have $ 1 - 2 \lambda_0 f^2 > 0 $ for all $f$.

If $ E_0 \le 0 $, then the right-hand side of \eqref{eq-GNLSE-caseII-2} is negative whenever $ f \neq 0 $, so it cannot equal the left-hand side; that is, the only $ f $ satisfying \eqref{eq-GNLSE-caseII-2} is identically zero. We therefore require $ E_0 > 0 $, i.e., 
    \begin{align}
    1 + \sigma b^2 - E > 0. \label{eq-GNLSE-caseII-constraint}
    \end{align}
Multiplying \eqref{eq-GNLSE-caseII-2} by $4f^2 / ( 1 - 2 \lambda_0 f^2 )$, we find
    \begin{align}
    (F')^2 = \frac{4 E_0 F^2 - 2 \kappa_0 F^3}{1 - 2 \lambda_0 F} := G(F), \label{eq-GNLSE-caseII-F}
    \end{align}
where we have defined $ F := f^2 $. 

Equation \eqref{eq-GNLSE-caseII-F} is invariant under translation and reflection (as is equation \eqref{eq-GNLSE-caseII}), meaning that if $ F(\xi) $ is a solution then so is $ F(const - \xi) $. Therefore we look for solutions which are reflectively symmetric about some point, which we set without loss of generality to $ \xi = 0 $. Then, $ F'(0) = 0 $ implies either $ F(0) = 0 $ or $ F(0) = 2 E_0 / \kappa_0 $. Suppose $ F(0) = 0 $; then, since $ F(\infty) = 0 $ and $F$ is smooth, there must exist some $\xi_0 \in (0,\infty)$ where $F'(\xi_0) = 0$ and $F(\xi_0) = 2 E_0 / \kappa_0 $. We therefore invoke the translational invariance again to set $\xi_0 = 0$ without loss of generality. Thus,
    \begin{align}
    F_0 := F(0) = \frac{2 E_0}{\kappa_0} > 0. \label{eq-GNLSE-caseII-F0}
    \end{align}
We see that $F(\xi)$ cannot exceed $F_0$, since if it did, then there is no greater value of $F$ at which $F'$ can vanish. That is to say, $ \max_{\xi \in \mathbb{R}} F (\xi) = F_0 $. It is easy to check that for all $F \in (0,F_0)$, we have $G(F) > 0$.

Since $F$ must vanish at infinity, we seek $F$ such that 
    \begin{align}
    F'(\xi) = - \textnormal{sgn} (\xi) \sqrt{G(F)} , \label{eq-GNLSE-caseII-3}
    \end{align}
where $ \textnormal{sgn} $ is the sign function. Using the change of variable 
    \begin{align}
    Z ( \xi ) = \textnormal{arsech} \sqrt{ F(\xi)/F_0 } ,
    \end{align}
and the fact that $ \sqrt{G(F)} = 2 \sqrt{E_0} F \sqrt{\smallfrac{1 - (F/F_0)}{1 - 2 \lambda_0 F}}$, we deduce
    \begin{align}
    Z'(\xi) = \frac{ \textnormal{sgn} (\xi) \sqrt{E_0} }{\sqrt{1 - 2 \lambda_0 F}} = \frac{ \textnormal{sgn} (\xi) \sqrt{E_0} }{\sqrt{1 - 2 \lambda_0 F_0~ \textnormal{sech}^2 Z}}. \label{eq-GNLSE-caseII-4}
    \end{align}
Integrating \eqref{eq-GNLSE-caseII-4} and invoking the vanishing boundary condition, we find
    \begin{align}
    \sqrt{E_0} ~| \xi | = \textnormal{\textnormal{arsinh}} \sqrt{ \frac{1 - \widetilde{F}}{\widetilde{F} - \mu^2 \widetilde{F}} } - \mu ~\textnormal{artanh} \sqrt{\frac{ 1 - \widetilde{F} }{(1/\mu^2) - \widetilde{F}}}, \label{eq-GNLSE-caseII-soln}
    \end{align}
where $ \mu = \sqrt{ 2 \lambda_0 F_0 } $, $ \widetilde{F} = F / F_0 $, and we have used the positivity of $Z$ to write $ \sinh Z = \sqrt{ (1/\widetilde{F}) - 1 } $.

\begin{figure}[t]
\centering
\includegraphics[width=0.75\textwidth]{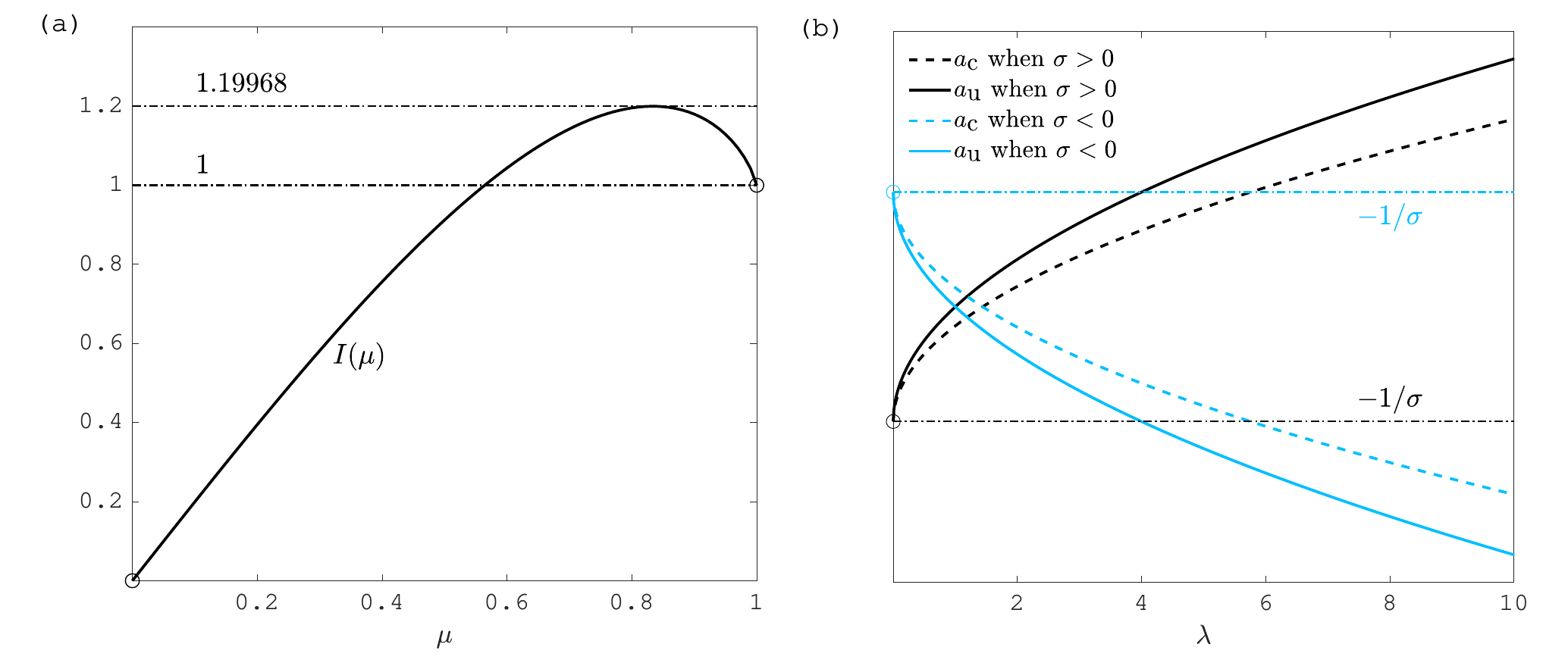}
\captionsetup{width=\textwidth}
\caption{The existence and uniqueness of $ \mu $ according to \eqref{eq-GNLSE-caseII-5}, depending on the values of $\kappa,\lambda,\sigma,a$. (a) $I(\mu)$ for $\mu \in (0,1)$. To every $I \in (0,1] \cup \{ \max I \approx 1.19968 \} $ there corresponds a unique $\mu$. To every $I \in (1,\max I)$ there corresponds two distinct $\mu$'s. (b) $a_{\textnormal{c}}(\lambda;\sigma)$ and $a_{\textnormal{u}}(\lambda;\sigma)$ are threshold values of $a$ such that $\sqrt{\lambda \kappa} / (1+\sigma a_{\textnormal{c}}) = \max I$ and $\sqrt{\lambda \kappa} / (1+\sigma a_{\textnormal{u}}) = 1$. The vertical scale is unspecified, since it is determined by the arbitrary $\kappa, \sigma $. If $\sigma > 0$ ($\sigma < 0$), then: $a \ge a_{\textnormal{u}}$ ($a \le a_{\textnormal{u}}$) or $a = a_{\textnormal{c}}$ implies unique $\mu$; $a_{\textnormal{c}} < a < a_{\textnormal{u}}$ ($a_{\textnormal{u}} < a < a_{\textnormal{c}}$) implies two distinct $\mu$'s; $-1/\sigma < a < a_{\textnormal{c}}$ ($a_{\textnormal{c}} < a < -1/\sigma$) implies no $\mu$. Given any $\kappa$ and $\sigma$, we always find $\lim_{\lambda \rightarrow 0} a_{\textnormal{c}} (\lambda) = \lim_{\lambda \rightarrow 0} a_{\textnormal{u}} (\lambda) = -1/\sigma$. Given any $\kappa$, we always find that the sign-reversal $\sigma \rightarrow -\sigma$ causes $a_{\textnormal{c}} \rightarrow - a_{\textnormal{c}}$ and $a_{\textnormal{u}} \rightarrow - a_{\textnormal{u}}$.}
\label{fig-1}
\end{figure}

The right-hand side of \eqref{eq-GNLSE-caseII-soln} is a strictly decreasing, unbounded function of $\widetilde{F}$ (setting $\widetilde{F} = 0$ makes the right-hand side blow up as required), so \eqref{eq-GNLSE-caseII-soln} uniquely determines $\widetilde{F}$ given any $\xi$. Re-introducing $x,y,t,\psi$ yields an exact solution to \eqref{eq-GNLSE}:
    \begin{subequations} \label{eq-GNLSE-caseII-soln-re}
    \begin{align}
    &\psi(x,y,t) = f(x+ay - c t) \exp [ i ( x+by - E t ) ] ,\quad \textnormal{where } \sigma b (1+a) = c - 2 \textnormal{ and} \\
    &\sqrt{\frac{ 1 + \sigma b^2 - E }{1 + \sigma a}} ~| x+ay - c t | = \textnormal{\textnormal{arsinh}} \sqrt{ \frac{1 - \widetilde{f}^2}{\widetilde{f}^2 - \mu^2 \widetilde{f}^2} } - \mu ~\textnormal{artanh} \sqrt{\frac{ 1 - \widetilde{f}^2 }{(1/\mu^2) - \widetilde{f}^2}},  \label{eq-GNLSE-caseII-soln-re2} \\
    &\textnormal{with } \mu = \sqrt{ \smallfrac{4 \lambda }{ \kappa ( 1 + \sigma a ) } \left( 1 + \sigma b^2 - E \right)} ,~ \widetilde{f} = f / f_0 \textnormal{ and } f_0^2 = f(0)^2 = \smallfrac{ 2 }{ \kappa } \left( 1 + \sigma b^2 - E \right) .
    \end{align}
    \end{subequations}
The solution \eqref{eq-GNLSE-caseII-soln-re} is parametrised by real constants $a,c,E$ satisfying the constraint [cf. \eqref{eq-GNLSE-caseII-constraint}]:
    \begin{align}
    H(a,c,E) := ( 1 + \sigma b^2 - E )(1+a)^2 \equiv (1 - E) (1+a)^2 + \frac{ (c-2)^2 }{ \sigma } \ge 0 . \label{eq-GNLSE-caseII-constraint2}
    \end{align}
The normalisation condition \eqref{eq-GNLSE-norm} imposes an extra constraint, which we derive as follows. Consider
    \begin{align}
    1 = \int_{-1/2}^{1/2} \int_{-\infty}^{\infty} | \psi (x,y,t) |^2 ~\d x ~\d y = \int_{-\infty}^{\infty} \frac{ f^2 }{ Z'(\xi) } \frac{\d x}{\d \xi} \d Z.
    \end{align}
Note that the $y$-integral is trivial because the $x$-domain is all of $\mathbb{R}$. Using the evenness of $f^2 = f_0^2 \textnormal{sech}^2 Z $, we further deduce
    \begin{align}
    1 = 2 \int_0^{\infty} \frac{ f^2}{ Z'(\xi > 0) } \d Z = 2 \int_0^{\infty} \frac{ ( f_0^2 \textnormal{sech}^2 Z ) \sqrt{ 1 - \mu^2 \textnormal{sech}^2 Z } }{ \sqrt{E_0} } \d Z,
    \end{align}
and hence
    \begin{align}
    \frac{ \sqrt{E_0} }{ 2 f_0^2 } = \frac{1}{2} + \frac{\left( 1 - \mu^2 \right) \textnormal{artanh} (\mu)}{2 \mu} . \label{eq-GNLSE-caseII-tr}
    \end{align}
Multiplying \eqref{eq-GNLSE-caseII-tr} by $ 2 \mu \equiv 2 \sqrt{ 2 \lambda_0 f_0^2 } $ yields a transcendental equation for $\mu$:
    \begin{align}
    \frac{ \sqrt{ \lambda \kappa } }{ 1 + \sigma a } = \mu + ( 1 - \mu^2 ) \textnormal{artanh} (\mu) := I(\mu) . \label{eq-GNLSE-caseII-5}
    \end{align}
By considering the graph of $I(\mu)$ for $\mu \in (0,1)$ [Figure \ref{fig-1}a], we conclude the following. Given any $\kappa,\lambda > 0$ and $\sigma > 0$ ($\sigma < 0$), the left-hand side of \eqref{eq-GNLSE-caseII-5} is a decreasing (increasing) function of $a > -1/\sigma$ ($a <-1/\sigma$), therefore: if $a$ is sufficiently large (small) then \eqref{eq-GNLSE-caseII-5} uniquely determines $\mu$; if $a$ is sufficiently small (large) then no value of $\mu$ satisfies \eqref{eq-GNLSE-caseII-5}; and if $a$ is moderate then \eqref{eq-GNLSE-caseII-5} determines two distinct values of $\mu$. The constant $a$ is therefore constrained by $a \ge a_{\textnormal{c}}(\lambda)$ if $\sigma > 0$ or $a \le a_{\textnormal{c}}(\lambda)$ if $\sigma < 0$, where $a_{\textnormal{c}}$ is such that $\sqrt{\lambda \kappa} / (1+\sigma a_{\textnormal{c}}) = \max I$ [Figure \ref{fig-1}b].

\subsection*{Case III. $ 1 + \sigma a < 0 $.}

If $ 1 + \sigma a < 0 $, then \eqref{eq-GNLSE-2} becomes $ E_1 f + ( 1 + 2 \lambda_1 f^2 ) f''  - \kappa_1 f^3 + 2 \lambda_1 (f')^2 f = 0 $, where $ E_1 = - \frac{1 + \sigma b^2 - E}{1 + \sigma a},  \lambda_1 = - \frac{\lambda}{1 + \sigma a} > 0,  \kappa_1 = - \frac{\kappa}{1 + \sigma a} > 0 $. Defining $ F = f^2 $, we find by analogy to case II that
    \begin{align}
    (F')^2 = \frac{- 4 E_1 F^2 + 2 \kappa_1 F^3}{1 + 2 \lambda_1 F} . \label{eq-GNLSE-caseIII-1}
    \end{align}
Using a similar argument as for \eqref{eq-GNLSE-caseII-F0}, we establish that $ F_0 := F(0) = 2 E_1 / \kappa_1 > 0 $. Now, for all $ F \in (F(\infty)=0,F_0) $, the numerator of the right-hand side of \eqref{eq-GNLSE-caseIII-1} is negative; since the denominator is always positive, we conclude that it is impossible for a non-trivial $F$ to satisfy \eqref{eq-GNLSE-caseIII-1} and therefore reject the case of $ 1 + \sigma a < 0 $ altogether.

\subsection*{Well-posedness and permissible parameter values}

Since \eqref{eq-GNLSE} is a quasilinear PDE, its well-posedness deserves careful consideration. Given $ \kappa $, and $ \sigma > 0 $ without loss of generality, the solution \eqref{eq-GNLSE-caseII-soln-re} is parametrised by $\lambda > 0$ and $a \ge a_{\textnormal{c}}(\lambda)$, with wave speed $c$ determined by [cf. \eqref{eq-GNLSE-caseII-constraint2}]:
    \begin{align}
    &c = c^{(\pm)} = \pm \sqrt{ \sigma [H + (E - 1)(1+a)^2] } + 2.
    \end{align}    
We denote by $ \psi^{(\pm)} $ the solution corresponding to $c = c^{(\pm)} $. Then, $ \psi^{(+)} $ and $ \psi^{(-)} $ are solitons with identical waveforms but distinct wave speeds, unless $c^{(+)} = c^{(-)}$. Since $ \psi^{(+)} (x,y,0) = \psi^{(-)} (x,y,0) $, we conclude that \eqref{eq-GNLSE} is well-posed \emph{only if} $H + (E-1)(1 + a)^2 = 0$, since any other case would give rise to non-unique solutions with identical initial data. Therefore, 
with $\sigma,\kappa$ and $\lambda$ prescribed for \eqref{eq-GNLSE}, we first choose $a$ such that $\mu$ exists, then find $ \mu $ through \eqref{eq-GNLSE-caseII-5}; invoking $c=2$, we find in turn:
    \begin{subequations} \label{eq-c-pm}
    \begin{align}
    H &= \mu^2 \kappa (1+\sigma a) (1+a)^2 / (4 \lambda) \ge 0, \label{eq-c-pm1} \\
    E &= 1 - \mu^2 \kappa (1+\sigma a) / (4 \lambda) < 1, \label{eq-c-pm2} \\ 
    b^2 &= \smallfrac{1}{\sigma} \left( \smallfrac{H}{(1+a)^2} + E - 1 \right) = 0 , \label{eq-c-pm3} \\
    f_0^2 &= \mu^2 ( 1 + \sigma a ) / (2 \lambda) > 0. \label{eq-c-pm4}
    \end{align}
    \end{subequations}
In particular, using \eqref{eq-c-pm3}, and the wave speed $ c = 2 $, we simplify the solution \eqref{eq-GNLSE-caseII-soln-re}:
    \begin{subequations} \label{eq-GNLSE-caseII-soln-simp}
    \begin{align}
    &\psi(x,y,t) = f(x+ay - 2 t) \exp [ i ( x - E t ) ] , \textnormal{ and} \\
    &\sqrt{\frac{ 1 - E }{1 + \sigma a}} ~| x+ay - 2 t | = \textnormal{\textnormal{arsinh}} \sqrt{ \frac{1 - \widetilde{f}^2}{\widetilde{f}^2 - \mu^2 \widetilde{f}^2} } - \mu ~\textnormal{artanh} \sqrt{\frac{ 1 - \widetilde{f}^2 }{(1/\mu^2) - \widetilde{f}^2}}, \textnormal{ with }  \widetilde{f} = f / f_0 . \label{eq-GNLSE-caseII-soln-simp2}
    \end{align}
    \end{subequations}
As we will discuss in Section \ref{sec-disc}, the physically relevant quantity is precisely $ | \psi |^2 \equiv f^2 $, which is now computable via \eqref{eq-GNLSE-caseII-soln-simp2}. The restriction of wave speed ensures non-singularity of this solution. Meanwhile, the solution is stable with respect to perturbations in parameter space if and only if $\lambda$ is sufficiently large; we demonstrate this point with the following example.
\begin{figure}[!t]
\centering
\includegraphics[width=0.5\textwidth]{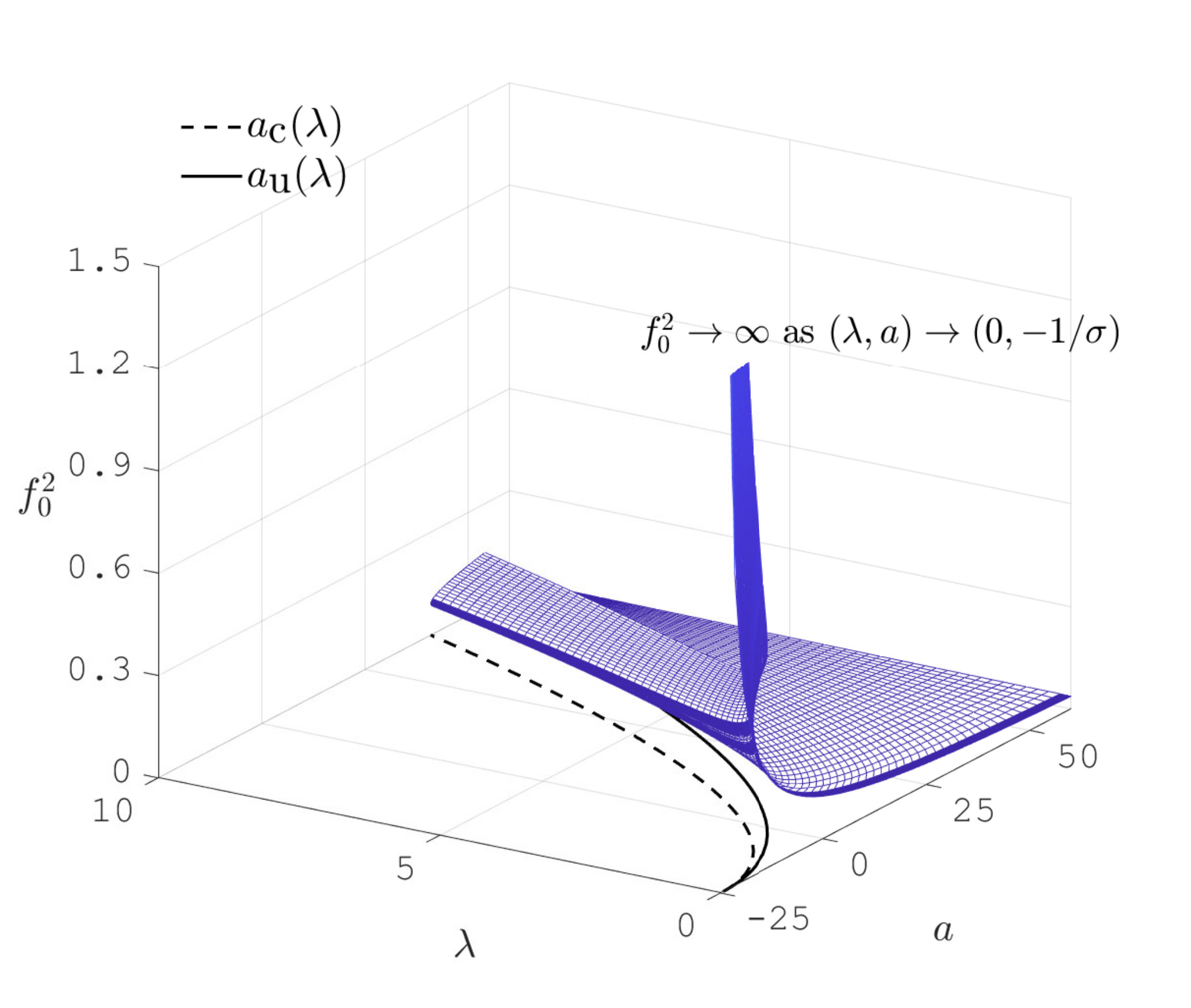}
\captionsetup{width=\textwidth}
\caption{The maximum squared amplitude $f_0^2$ as a function of the anisotropy parameter $\lambda$ and waveform constant $a$, with $\sigma = 0.04, \kappa = 1$ and $E = 1$ fixed. The $f_0^2$ surface is folded sharply along the curve $a = a_{\textnormal{c}}(\lambda)$, so that to each $a \in (a_{\textnormal{c}}(\lambda) , a_{\textnormal{u}}(\lambda))$ there corresponds two distinct values of $f_0^2$, and for $ a = a_{\textnormal{c}} (\lambda) $ or $ a \ge a_{\textnormal{u}}(\lambda) $ there is a unique $f_0^2$. If $(\lambda,a) \rightarrow (0,-1/\sigma)$ along any curve $a(\lambda)$ where $\sqrt{\lambda \kappa} / (1+\sigma a)$ is constant, e.g.~$a = a_{\textnormal{c}}$ or $a_{\textnormal{u}}$, then $f_0^2$ blows up. Since all such $a(\lambda)$ curves unify as $\lambda \rightarrow 0$, we have $f_0^2 \rightarrow \infty$ as $(\lambda,a) \rightarrow (0,-1/\sigma)$.}
\label{fig-2}
\end{figure}
\begin{figure}[!t]
\centering
\includegraphics[width=\textwidth]{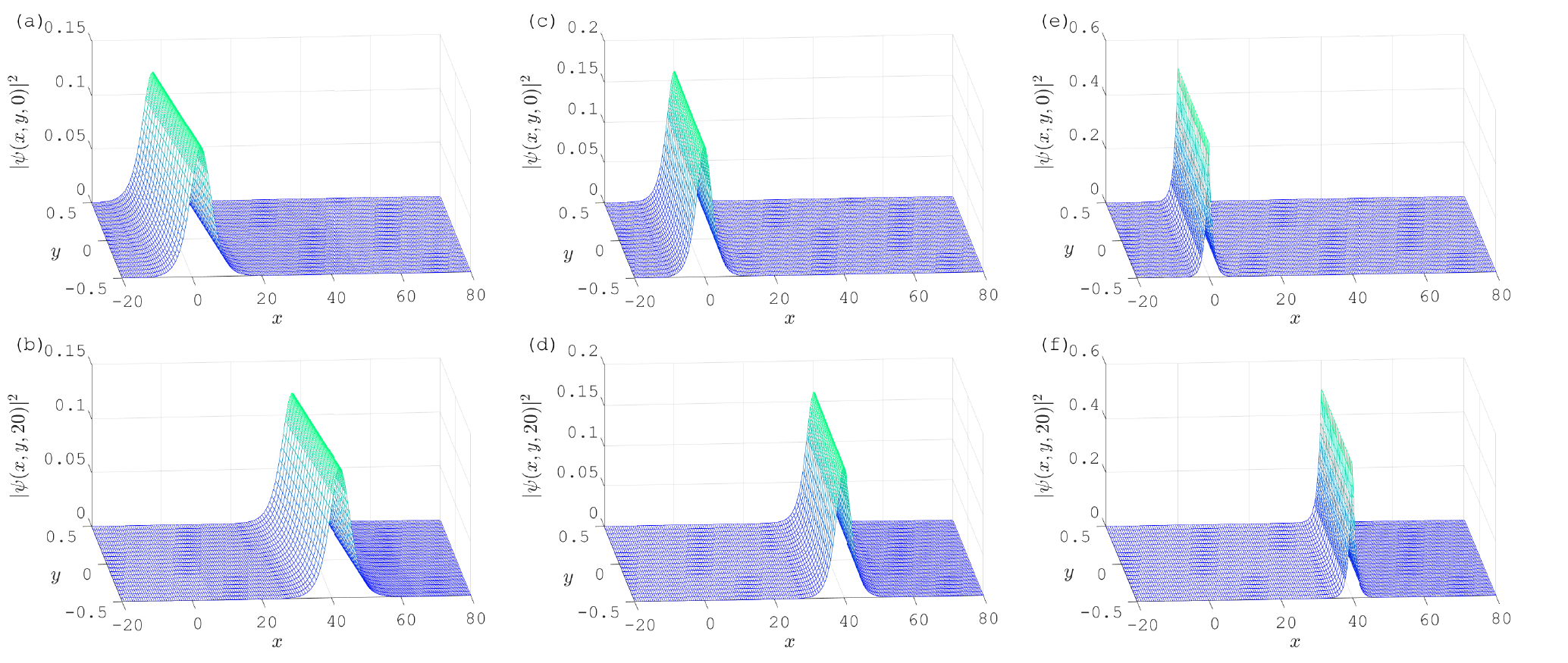}
\captionsetup{width=\textwidth}
\caption{Propagation of the solitary wave $ | \psi |^2 $ which solves equation \eqref{eq-GNLSE} with $\sigma = 0.04, \kappa = 1$ and $\lambda = 1$. The solution is parametrised by the waveform constant $a$. (a,b) $a=5$, to which there corresponds a unique waveform. (c,d,e,f) $a=-0.2$, to which there correspond two distinct waveforms; one is represented in (c,d), the other in (e,f). (a,c,e): $ |\psi|^2 $ at $t=0$. (b,d,f): $ |\psi|^2 $ at $t=20$.}
\label{fig-3}
\end{figure}

We consider $ \sigma = 0.04 $ and $ \kappa = 1 $ for the purpose of this illustration; the solution is then parametrised by $\lambda > 0$ and $a \ge a_{\textnormal{c}}(\lambda)$ [Figure \ref{fig-2}]. In the $ a \ge a_{\textnormal{u}}(\lambda) $ region of the $(\lambda,a)$ parameter space, which we call the \textit{uni-waveform} region, $f_0^2(\lambda,a)$ is unique and barely affected by varying $\lambda$; whereas in the $ a_{\textnormal{c}}(\lambda) < a < a_{\textnormal{u}}(\lambda) $ region, which we call the \textit{bi-waveform} region, $f_0^2$ is bi-valued and one branch is more sensitive to variations in $a$ and $\lambda$ than the other branch. As $(\lambda,a) \rightarrow (0,-1/\sigma)$, $f_0^2$ blows up, meaning that the solitary wave asymptotically approaches the delta distribution. Mathematically, the blow-up of $f_0^2$ is predictable from \eqref{eq-c-pm}: if $\kappa$ and $\sqrt{\lambda \kappa} / (1+\sigma a)$ are both held constant, then as $ (\lambda,a) \rightarrow (0,-1/\sigma) $, $\mu$ remains fixed at a finite value, hence $f_0^2 \sim 1/\sqrt{\lambda}$. Since all the $a(\lambda)$ curves hold $\sqrt{\lambda \kappa} / (1+\sigma a)$ constant, and in the limit $\lambda \rightarrow 0$ they unify into $(\lambda,a) \rightarrow (0,-1/\sigma)$, we conclude that $f_0^2 \rightarrow \infty$ as $(\lambda,a) \rightarrow (0,-1/\sigma)$. The blow-up of $f_0^2$ for small $\lambda$ indicates instability of the solution in this region of parameter space. 

Some representations of the $|\psi|^2$ waves can be found in Figure \ref{fig-3}, with prescribed parameters $\sigma = 0.04, \kappa = 1$ and $\lambda = 1$. For $a = 5$, there is a unique $\mu$ and hence a unique waveform, initially maximised along the line $x+5y = 0$ [Figure \ref{fig-3}a]. For $a = -0.2$, two distinct waveforms exist: both maximised along the line $x - 0.2y = 0$ initially, but one of them has maximum amplitude $ \approx 0.16 $ [Figure \ref{fig-3}c], whereas the other has maximum amplitude $ \approx 0.49 $ [Figure \ref{fig-3}e]. 

    \section{Origins in condensed matter physics} \label{sec-disc}
    
To interpret the solution \eqref{eq-GNLSE-caseII-soln-simp}, we begin by deriving \eqref{eq-GNLSE} as a model of the dynamics of a quantum mechanical exciton when coupled to a lattice. The theoretical principles of exciton-lattice interaction were founded in the 1930s, and have been important in condensed matter physics for their relevance to superconductivity, conducting polymers and bioelectronics \cite{landau1933electron,frohlich1950theory,holstein1959studies,su1979solitons,brizhik2019long}. The method of deriving the NLSE or its variants by taking a exciton-lattice system to a continuum limit was first presented by Davydov \cite{davydov1979solitons}, and has since been employed in many studies; details can be found in some excellent reviews \cite{heeger1988solitons,scott1992davydov}. Since the theory is well established, we present only a brief outline of key steps.

We consider a 2D rectangular lattice whose number of nodes is large in one direction, say $x$, and let the lattice undergo collective stretching oscillations along that direction. An extra exciton, such as an electron, interacts with the lattice and is assumed to undergo no spin-flip. The exciton-lattice system is modelled with a second-quantised Fr\"{o}hlich-Holstein Hamiltonian:
    \begin{align}
	\widehat{H} &= J_0 \sum_{ k = 0 }^{ N_2 } \sum_{ j = 0 }^{ N_1 } \widehat{A}^{\dagger}_{ j , k } \widehat{A}_{ j , k } - J_1 \sum_{ k = 0 }^{ N_2 } \sum_{ j = 0 }^{ N_1 - 1 } \Big( \widehat{A}^{\dagger}_{ j + 1 , k } \widehat{A}_{ j , k } + \widehat{A}^{\dagger}_{ j , k } \widehat{A}_{ j + 1 , k } \Big) \nonumber \\
	& \quad - J_2 \sum_{ k = 0 }^{ N_2 - 1 } \sum_{ j = 0 }^{ N_1 } \Big( \widehat{A}^{\dagger}_{ j , k + 1 } \widehat{A}_{ j , k } + \widehat{A}^{\dagger}_{ j , k } \widehat{A}_{ j , k + 1 } \Big) + \sum_{ k = 0 }^{ N_2 } \sum_{ j = 0 }^{ N_1 } \frac{ \widehat{P}_{ j , k }^2 }{ 2 M } + \sum_{ k = 0 }^{ N_2 } \sum_{ j = 0 }^{ N_1 - 1 } \frac{ K }{ 2 } \left( \widehat{Q}_{ j + 1 , k } - \widehat{Q}_{ j , k } \right)^2 \nonumber \\
	& \quad + \sum_{ k = 0 }^{ N_2 } \left[ \chi_2 \left( \widehat{Q}_{ 1 , k } - \widehat{Q}_{ 0 , k } \right) \widehat{A}^{\dagger}_{ 0 , k } \widehat{A}_{ 0 , k } + \chi_1 \left( \widehat{Q}_{ N_1 , k } - \widehat{Q}_{ N_1 - 1 , k } \right) \widehat{A}^{\dagger}_{ N_1 , k } \widehat{A}_{ N_1 , k } \right] \nonumber  \\
	& \quad + \sum_{ k = 0 }^{ N_2 } \sum_{ j = 1 }^{ N_1 - 1 } \left[ \chi_2 \left( \widehat{Q}_{ j + 1 , k } - \widehat{Q}_{ j , k } \right) + \chi_1 \left( \widehat{Q}_{ j , k } - \widehat{Q}_{ j - 1 , k } \right) \right] \widehat{A}^{\dagger}_{ j , k } \widehat{A}_{ j , k }, \label{eq-ham}
	\end{align}
where $N_1+1$ and $N_2+1$ are the number of nodes in the $x$ and $y$ directions of the lattice, respectively; the remaining notations are explained as follows. The first three terms of the right-hand side of \eqref{eq-ham} constitute the standard tight-binding exciton model, with $J_0$ being the exciton site energy and $J_{1,2}$ the exciton transfer integrals; $\widehat{A}^{\dagger}_{j,k}$ and $\widehat{A}_{j,k}$ are the exciton creation and annihilation operators at the $(j,k)$ node, respectively. The equilibrium distances between lattice nodes are implicitly built into the constants $J_{1,2}$, and we assume $J_1 > 0$; the sign of $J_2 \neq 0$ can vary depending on the type of physical system \cite{scott1982dynamics,zolotaryuk1998polaron}. The next two terms constitute a masses-and-springs model of the lattice, with $M$ being the node mass, $K$ the spring constant, $\widehat{Q}_{j,k}$ the operator for node displacement from equilibrium, and $\widehat{P}_{j,k}$ being the momentum conjugate to $\widehat{Q}_{j,k}$. The remaining terms on the right-hand side of \eqref{eq-ham} model the exciton-lattice interaction, with coupling constants $\chi_{1,2} \ge 0$, not both zero, representing the interaction strengths between a localised exciton and lattice distortions to the small-$j$ and large-$j$ side, respectively. As we shall see, it is precisely this spatial anisotropy in exciton-lattice coupling that leads to the novel $ \partial_{xx} ( | \psi |^2 ) \psi $ term in \eqref{eq-GNLSE}; and it has been argued that an anisotropy of this type is important in modelling molecular-biological systems such as the $\alpha$-helix \cite{Luo2017,georgiev2019quantum}.

Assuming a disentangled exciton-lattice system where the exciton is in a Bloch state and the lattice in a Glauber state: 
    \begin{align}
	\ket{ \Psi ( t ) } = \sum_{ k = 0 }^{ N_2 } \sum_{ j = 0 }^{ N_1 } \phi_{ j , k } ( t ) \widehat{A}^{\dagger}_{ j , k } \exp \left( \frac{ i }{ \hbar } \sum_{ k' = 0 }^{ N_2 } \sum_{ j' = 0 }^{ N_1 } \left( \mathcal{P}_{ j' , k' } ( t ) \widehat{Q}_{ j' , k' } - \mathcal{Q}_{ j' , k' } ( t ) \widehat{P}_{ j' , k' } \right) \right) \ket{ 0_\textnormal{e} } \ket{ 0_\textnormal{p} } , \label{eq-state}
	\end{align}
where $ \ket{ 0_\textnormal{e} } $ and $\ket{ 0_\textnormal{p}}$ are the exciton and lattice vacua respectively, we follow the standard Hamiltonian procedure to derive the equations governing the dynamics of the coefficients $ \phi_{j,k}, \mathcal{P}_{j,k} $ and $ \mathcal{Q}_{j,k} $, for the \textit{interior points} $j = 1,2, \dots N_1 - 1, k = 1,2, \dots N_2 - 1$:
    \begin{subequations} \label{eq-disc}
    \begin{align}
	i \frac{ \d \phi_{ j , k } }{ \d t } &= \phi_{ j , k } \left[ \frac{ J_0 }{ J_1 } + \Omega + \frac{ \alpha }{ 2 } ( 1 + \beta ) ( q_{ j + 1 , k } - q_{ j , k } ) + \frac{ \alpha }{ 2 } ( 1 - \beta ) ( q_{ j , k } - q_{ j - 1 , k } ) \right] \nonumber \\
	&\quad - [ \phi_{ j - 1 , k } + \phi_{ j + 1 , k } ] - \rho [ \phi_{ j , k - 1 } + \phi_{ j , k + 1 } ] , \label{eq-disc-phi} \\
	\frac{ \d^2 q_{ j , k } }{ \d t^2 } &= \gamma^2 \left[ ( q_{ j + 1 , k } - 2 q_{ j , k } + q_{ j - 1 , k } ) \right] + \frac{ \alpha }{ 2 } ( 1 - \beta ) \left[ \abs{ \phi_{ j + 1 , k } }^2 - \abs{ \phi_{ j , k } }^2 \right] + \frac{ \alpha }{ 2 } ( 1 + \beta ) \left[ \abs{ \phi_{ j , k } }^2 - \abs{ \phi_{ j - 1 , k } }^2 \right]. \label{eq-disc-q}
	\end{align}
	\end{subequations}
In deriving \eqref{eq-disc}, we have scaled time by $ \hbar / J_1 $, and scaled length by $ \hbar / \sqrt{M J_1} $ so that $ q_{j,k} $ is the non-dimensionalised $ \mathcal{Q}_{j,k} \equiv \braket{\Psi | \widehat{Q}_{j,k} | \Psi} $; we have also defined the dimensionless constants $ \alpha = ( \chi_2 + \chi_1 ) \hbar / \sqrt{M J_1^3} > 0 , \beta = (\chi_2 - \chi_1)/( \chi_2 + \chi_1 ), \rho = J_2 / J_1 \neq 0 , \gamma = \sqrt{ K \hbar^2 / ( M J_1^2 ) } > 0 $, and the dimensionless lattice energy which we assume to be constant: 
    \begin{align}
    \Omega = \frac{ \Braket{ \Psi | \sum_k \sum_j \big( \frac{ \widehat{P}_{ j , k }^2 }{ 2 M } + \frac{ K }{ 2 } ( \widehat{Q}_{ j + 1 , k } - \widehat{Q}_{ j , k } )^2 \big) | \Psi } }{ J_1 } .
    \end{align}
Note that $\beta$ encodes the anisotropy of the exciton-lattice interaction, and should take values in $[-1,1]$. However, as we will soon discover, setting $ \beta = 0 $ simply reduces the system to the standard NLSE with a cubic nonlinearity (equivalent to setting $ \lambda = 0 $ in \eqref{eq-GNLSE}). The standard NLSE and its solutions are so well known that we do not consider it here. Furthermore, we assume without loss of generality that $ \chi_2 > \chi_1 $, so that $ \beta \in (0,1] $.

Now we invoke the continuum approximation by introducing smooth functions $ \psi , u $ of the continuous variables $(x,y,t)$ such that $ \phi_{j,k} (t) = \psi (j,k,t) \exp [ - i t (J_0/J_1 + \Omega - 2 - 2 \rho) ]$ and $ q_{j,k} (t) = u(j,k,t) $. Approximating finite differences by derivatives up to the second order \cite{davydov1979solitons}, we find
    \begin{subequations} \label{eq-cont}
    \begin{align}
	i \partial_t \psi &= \psi \times \left( \alpha \partial_x u + \frac{ \alpha \beta }{2} \partial_{xx} u \right) - \partial_{xx} \psi - \rho \partial_{yy} \psi , \label{eq-cont-psi} \\
	\partial_{tt} u &= \gamma^2 \partial_{xx} u + \alpha \partial_x | \psi |^2 - \frac{ \alpha \beta }{ 2 } \partial_{xx} | \psi |^2. \label{eq-cont-u}
	\end{align}
	\end{subequations}
Differentiating \eqref{eq-cont-u} yields $ \partial_{tt} (\partial_x u) - \gamma^2 \partial_{xx} ( \partial_x u ) = \partial_{xx} ( \alpha | \psi |^2 - \smallfrac{ \alpha \beta }{ 2 } \partial_{x} | \psi |^2 )$. We consider a travelling wave ansatz: $ \partial_x u (x,y,t) = v (x-wt) $ for some function $v$ and constant $w$, which leads to $ \partial_{xx} [ (w^2 - \gamma^2 ) v ] = \partial_{xx} ( \alpha | \psi |^2 - \smallfrac{ \alpha \beta }{2} \partial_x | \psi |^2 )$ and therefore enables us to take $ v = ( \alpha | \psi |^2 - \smallfrac{ \alpha \beta }{2} \partial_x | \psi |^2 ) / (w^2 - \gamma^2) $. Substituting this expression for $ \partial_x u $ in \eqref{eq-cont-psi}, we deduce
    \begin{align}
	i \partial_t \psi &= \psi \times \left( \frac{\alpha^2}{w^2 - \gamma^2} | \psi |^2 - \frac{ \alpha^2 \beta^2 }{4 (w^2 - \gamma^2)} \partial_{xx} | \psi |^2 \right) - \partial_{xx} \psi - \rho \partial_{yy} \psi \label{eq-cont-2}. 
    \end{align}    
Equation \eqref{eq-cont-2} holds over the interior of a finite, rectangular spatial domain, but we scale the $y$ variable so that $ y \in (-\smallfrac{1}{2},\smallfrac{1}{2}) $; and since the lattice is large in the $x$ direction, we let $ x \in \mathbb{R} $. We require that the speed of the $ \partial_x u $ wave is sufficiently small so that $ w^2 < \gamma^2 $ (which we will soon justify). Rearranging \eqref{eq-cont-2}, defining $ \kappa = \alpha^2 / ( \gamma^2 - w^2 ) > 0 $, $ \lambda = \alpha^2 \beta^2 / (4 (\gamma^2 - w^2) ) > 0 $, and combining $\rho$ with the appropriate $y$-scaling into a constant $ \sigma \neq 0 $, we obtain exactly the equation \eqref{eq-GNLSE}.

The normalisation condition \eqref{eq-GNLSE-norm} is imposed because the quantum state $ \ket{ \Psi } $ as per \eqref{eq-state} must be normalised. In the limit $ \beta \rightarrow 0 $, we recover the standard NLSE; that is to say, the exciton-lattice system with spatially isotropic coupling is modelled in the continuum limit by the standard NLSE, as is well known \cite{davydov1979solitons}. We have required $ w^2 < \gamma^2 $ to ensure $\kappa > 0$ and so that, in the limit $\beta \rightarrow 0$, we recover the NLSE that admits \textit{bright} solitons, rather than dark ones, which may not be normalisable. In a realistic system, we expect $ | \rho | \sim 1 $ and hence $ |\sigma| $ to be comparable to the squared reciprocal of the number of lattice nodes in the $y$ direction. 

In light of the discussions above, we say that the soliton solution \eqref{eq-GNLSE-caseII-soln-simp} represents the lossless propagation of an exciton wavefunction in a planar lattice with a large aspect ratio, with possible applicability to photonic crystal circuits and Bragg gratings \cite{lonvcar2000waveguiding,gnan2006modelling}. The squared amplitude, $|\psi|^2 \equiv f^2$, is the probability distribution of exciton location. The squared amplitude is a travelling wave with constant speed; it is maximised along a straight line, representing the most probable location of the exciton, and the line advances through the lattice. The parameters $\kappa, \lambda, \sigma$ encode the exciton-lattice coupling strength, the anisotropy of the coupling in the $+x$ versus $-x$ directions, and the anisotropy of exciton hopping energy in the $y$ versus $x$ directions, respectively. If we choose $w$ such that the velocity of the lattice distortion matches the propagation velocity of the exciton, and if the exciton is an electron, then the quasi-particle comprising lattice distortion and electron becomes what is well known in condensed matter physics as a \textit{polaron}.

    \section{Conclusions}

In this study, we have presented a modified nonlinear Schr\"{o}dinger equation in two spatial dimensions, featuring a nonlinear term of the form $ \partial_{xx} ( |\psi|^2) \psi $. We have constructed an exact soliton solution analytically, and characterised its properties, which depend on parameters $ (\sigma,\kappa,\lambda) $ in the equation and a key parameter in the travelling wave ansatz: a waveform constant $a$ determining the gradient of the line of maximum $ | \psi |^2 $. In particular, the waveform constant can only take values in a range $a \ge a_{\textnormal{c}}$ if $\sigma > 0$ or $ a \le a_{\textnormal{c}} $ if $\sigma < 0$, where the threshold $a_{\textnormal{c}}$ depends on $(\sigma,\kappa,\lambda)$ in ways which we have detailed. Another $a$-value threshold, $a_{\textnormal{u}}$, marks the bifurcation between having a unique waveform for each value of $a$, and having two distinct waveforms corresponding to each $a$. The dependence of $a_{\textnormal{u}}$ on $ (\sigma,\kappa,\lambda) $ has been established. The region of parameter space that lies strictly between $a_{\textnormal{c}}$ and $a_{\textnormal{u}}$ is the bi-waveform region. This waveform duality is a novel phenomenon among modified NLSE systems that admit soliton solutions. Crucially, we have found that regardless of the waveform, the wave speed must be $c=2$ in order to ensure the uniqueness of the soliton solution. 

From condensed matter theory, we have derived the equation as a model for the propagation of a quantum exciton through a plane lattice to which the exciton is coupled. The lattice is required to be much larger in one direction, say $x$, than in the other, and undergoing collective oscillations (e.g.~by hydrogen bond stretching) in that $x$ direction. The novelty of the model manifests as an anisotropy parameter $ \lambda > 0 $ which encodes the extent to which the exciton-lattice interaction is spatially anisotropic, with the limit $ \lambda \rightarrow 0 $ reducing the model to the standard nonlinear Schr\"{o}dinger equation. We have found that the solution is stable in regions of the parameter space with sufficiently large $\lambda$, representing exciton-lattice systems with highly anisotropic coupling. Furthermore, $ \psi $ is the exciton position wavefunction, and therefore the soliton solution represents the lossless transport of the exciton. If one wishes to consider higher-dimensional slender lattices, then the exciton-lattice system will be suitably modelled by the natural high-dimensional extension of the equation considered here, and the same method of solution will apply. Indeed, in three spatial dimensions this equation becomes a generalisation of the Gross-Pitaevskii model for Bose-Einstein condensates \cite{garcia2003quasi}.

\subsection*{Statement of author contribution(s)}
This work was carried out in full by the sole author.

\subsection*{Declaration}
The author declares no conflict of interest.

\subsection*{Acknowledgement}
The author thanks the University of Birmingham, UK for Fellowship funding.

\end{document}